\documentclass[12pt]{article}

\pdfoutput=1

\usepackage[english]{babel}
\usepackage{fancyhdr}
\usepackage{amsmath}
\usepackage{amssymb}
\usepackage{amsfonts}
\usepackage{psfrag}
\usepackage[applemac]{inputenc}
\usepackage{graphicx,wrapfig}
\usepackage{pstricks}

\usepackage{slashed}            
\usepackage{subfig}             
\usepackage{xspace}				
\usepackage[sort&compress,square,comma,numbers]{natbib}
\usepackage{float}
\usepackage{dcolumn}
\usepackage{bm}
\usepackage{epstopdf}
\usepackage{feynmp}

\usepackage[
	colorlinks=true,
	citecolor=black,
	linkcolor=black,
	urlcolor=blue,
	hypertexnames=false]{hyperref}
	
\newcommand{\arXiv}[2]{\href{http://arxiv.org/pdf/#1}{{\tt #2/#1}}}
\newcommand{\arXivold}[1]{\href{http://arxiv.org/pdf/#1}{{\tt #1}}}

\DeclareFontFamily{OT1}{pzc}{}
\DeclareFontShape{OT1}{pzc}{m}{it}{<-> s * [1.100] pzcmi7t}{}
\DeclareMathAlphabet{\mathpzc}{OT1}{pzc}{m}{it}

\newcommand{\beq}{\begin{eqnarray}}
\newcommand{\eeq}{\end{eqnarray}}
\newcommand{\bea}{\begin{eqnarray}}
\newcommand{\eea}{\end{eqnarray}}
\newcommand{\bag}{\begin{align}}
\newcommand{\eag}{\end{align}}

\newcommand{\TeV}{\,\mathrm{TeV}}

\begin{document}

\baselineskip=18pt

\setcounter{footnote}{0}
\setcounter{figure}{0}
\setcounter{table}{0}


\begin{titlepage}

\begin{center}
  \begin{LARGE}
    \begin{bf} A Diphoton Resonance from Bulk RS
   \end{bf}
  \end{LARGE}
\end{center}
\vspace{0.1cm}
\begin{center}
\begin{large}
{\bf Csaba Cs\'aki$^a$ and Lisa Randall$^b$ \\}
\end{large}
  \vspace{0.5cm}
  \begin{it}

\begin{small}
$^{(a)}$Department of Physics, LEPP, Cornell University, Ithaca, NY 14853, USA
\vspace{0.2cm}\\
$^{(b)}$Department of Physics, Harvard University, Cambridge, MA  02138 USA
\vspace{0.2cm}\\

\end{small}

\end{it}
\vspace{.5cm}

{\tt csaki@cornell.edu, randall@physics.harvard.edu}

\end{center}

\vspace*{0.5cm}

\begin{abstract}
\medskip
\noindent

\end{abstract}
Recent LHC data hinted at a 750 GeV mass resonance that decays into two photons. A significant feature of this resonance is that its decays to any other Standard Model particles would be too low to be detected so far. 
Such a state has a compelling explanation in terms of a  scalar or a pseudoscalar that is strongly coupled to vector states charged under the Standard Model gauge groups.   Such a scenario is readily accommodated  in bulk RS with a scalar localized in the bulk away from but close to the Higgs. Turning this around, we argue that a good way to find the elusive bulk RS model might
be the search for a resonance with prominent couplings to gauge bosons.

\bigskip

\end{titlepage}

\section{Introduction} 

This paper is dedicated to the memory of Yoichiro Nambu, who made immense contributions to particle phenomenology and in particular to the physics of strong dynamics. A preliminary version of this work was presented at~\cite{LisaTalk}.

Recent data from the ATLAS~\cite{ATLAS} and CMS~\cite{CMS} experiments hint at a 750 GeV resonance that decays into two photons. The event rate, the absence of other signatures, or of a significant signal in the lower-energy Run 1 of the LHC~\cite{Aad:2015mna,CMSgg8}, argues for a scalar or pseudoscalar coupled reasonably strongly to vector-like fermions charged under the Standard Model gauge group~\cite{Nomura,Rattazzi,Michele,Patrick,Gilad,Tomer}. While SM charged scalar interactions  instead of that of vector fermions are also possible, explaining their low mass would generally require additional  assumptions.

The mass of the fermions must be at least of order TeV in order to explain the lack of direct detection. To achieve the necessary production rates, the coupling of these fermions to the scalar would have to be relatively large, hinting at a strongly coupled theory. Even if there is large multiplicity, large running would argue for some sort of strong dynamics as well.

In this paper we explore a class of models in which a scalar resonance is sequestered from the IR brane in a fifth warped spatial dimension. Such models~\cite{RS} automatically have several features
consistent with the required constraints.

\begin{itemize}
\item Bulk RS models have the potential to explain flavor when fermions are in the bulk. Bulk fermions guarantee a large number of 
vector fermion states--the KK fermion modes-- charged under the Standard Model and distributed throughout the bulk.
\item With KK masses of order the experimental limit, the required Yukawas of the scalar to the five-dimenisonal fermions can be of order unity.
\item The scalar in the bulk couples to vector KK modes located in the same region. Because the SM is chiral, projection operators guarantee that
one chirality of the KK modes vanishes in the IR so ONLY  scalars with substantial support away from the IR brane  have the necessary interactions.
\item All Yukawas can be of order unity, also consistent with the absence of observed decays to the weak gauge bosons but potentially leading to additional observable signals.
\item A bulk theory with a third brane permits the possibility of lowering KK masses, thereby enhancing the decays to photons relative to Higgses.

\end{itemize}

The parameters of such a model are the bulk KK mass, the location of the additional scalar, and the Yukawas. Although in principle this is a large number of parameters,
we emphasize that all can be of order unity consistent with experimental constraints. The flexibility in Yukawas does however make it harder to predict precise ratios of the various
partial decay widths. From this perspective, one of the best experimental signatures of this model would be that the scalar has the right parity properties, as no other compelling strongly
interacting theory involving a single scalar exists.

 The presence of such scalars would actually be quite generic for bulk RS models. In the CFT picture they would correspond to additional states that become massive before the confining dynamics is reached, perhaps being responsible for the onset of confinement itself. Alternatively, in overlapping work, \cite{Matthias} suggested a scalar that corresponds to the same strong dynamics responsible for the IR brane, but which has a boundary condition that suppresses its support on the IR brane itself, leading to results similar to those we present.


In either case, the induced coupling to SM gauge bosons are generic, and would perhaps be  among the first observable signals of such models. In fact, it might be possible that such diboson resonances will be the only signals of bulk RS models, if the SM KK modes turn out to be just too heavy to be observed at the LHC.

\section{The Setup}

We consider a warped 5 dimensional RS2 theory~\cite{RS}, with all SM fields in the bulk~\cite{RSbulk}, and a Higgs field sharply peaked on the IR brane. We use the conformally flat form of the AdS$_5$ metric
\begin{equation}
ds^2 = \left(\frac{R}{z}\right)^2 (dx^2-dz^2)
\end{equation}
with the UV brane placed at $z=R$ and the IR brane at $z=R'$. $R$ is also the AdS curvature $R\sim M_{Pl}^{-1}$ and $R' \sim 1/$TeV. 

The key additional ingredient will be the assumption of an additional brane at $z = z_0 R'$, where $z_0<1$, $z_0 = {\cal O}(1)$. The additional scalar $S$ (assumed to be a singlet under the SM gauge interactions) will be localized on this brane. We will comment on the interpretation and the effect of this third brane in Sec.~\ref{sec:CFT}. This scalar will be identified with the 750 GeV resonance decaying to diphotons.  This scalar is assumed to have a Yukawa interaction with the bulk fermion fields. Since $S$ is a SM  singlet, it will have Yukawa couplings between the LH and the RH modes of the same 5D fermions:
\begin{equation}
\int d^4 x \sqrt{g^{ind}} Y_i R \tilde{S} \bar{\Psi}_i\Psi_i\ \ \to \ \ \int d^4 x \left[ y^{(eff)}_i S \chi_i^{(1)}  \psi_i^{(1)} + h.c. \right]
\label{eq:SYukawa}
\end{equation} 
where $\Psi$ is a bulk fermion, whose KK decomposition is given by~\cite{fermions} \begin{equation}
\Psi = \sum_n \left( g_n(z) \chi_n (x) + f_n (z) \bar{\psi}_n (x) \right)
\end{equation}
with $\chi$ always denoting a LH 4D 2 component fermion and $\bar{\psi}$ a RH 2 component fermion $\Psi = (\chi , \bar{\psi})$. There is a separate bulk field for each SM fermion: for the LH SM fields we denote the corresponding bulk field by $\Psi_L$, which will yield a zero mode in $\chi_L$, while for the RH SM fields one has a bulk multiplet $\Psi_R$, with a zero mode in $\psi_R$. Each of these multiplets has a bulk mass given by $\frac{c_{L,R}}{R}$, where the bulk mass parameter $c_{L,R}$ controls the shape of the zero mode wave functions, and also plays a role in the shape of the KK modes.  LH zero modes are UV localized for $c_L>0.5$, while RH zero modes for $c_R<-0.5$. The bulk equations of motion for the fermions imply the following wave functions for the n$^{th}$ KK mode: 

\begin{eqnarray}
g_n (z) = \left( \frac{z}{R}\right)^\frac{5}{2} \left (A_n J_{c+\frac{1}{2}} ( m_n z) + B_n J_{-c-\frac{1}{2}} ( m_n z) \right) \\
f_n (z) = \left( \frac{z}{R}\right)^\frac{5}{2} \left (A_n J_{c-\frac{1}{2}} ( m_n z) - B_n J_{-c+\frac{1}{2}} ( m_n z) \right) 
\end{eqnarray}

The boundary conditions of fermions corresponding to LH (or RH)  SM fields (in the absence of the Higgs VEV) is 
\begin{equation}
\psi_L|_R = \psi_L|_{R'}= 0, \ \ \ \chi_R|_R = \chi_R|_{R'}= 0\ .
\end{equation}
The lowest KK mode will be given by $x_1/R'$ with $x_1 \approx 2.45$, for LH fermions $J_{1/2-c_L}(x_1)=0$.

One of the chiralities participating in the Yukawa interactions in (\ref{eq:SYukawa}) is vanishing on the IR brane - thus such Yukawa coupling can actually only exist at a position somewhat away from the IR brane. 

Using the KK decomposition of the fermions we have numerically caluclated the effective suppression of the effective Yukawa coupling between $S$ and the first KK mode of a bulk fermion as a function of the position of the S-brane, and also varying the bulk mass parameter $c$. The coupling in the immediate vicinity of the IR brane is suppressed since one of the chiralities vanishes there, then it peaks around $z_0 \sim 0.7$, and is again suppressed towards the UV  brane. Overlaps of around $0.8$ can be achieved. 

 Note that the phenomenology of the proposed resonance argues for a near-IR scalar. Were we to allow for stronger Yukawa couplings, the interactions could be generated by more massive KK modes. When the dominant contributions comes from the lowest-lying KK modes, the scalar has to be localized near the IR brane, where the light KK modes have the greatest support. In this case, the scalar, like the unphysical chirality KK modes, might be near-IR localized because of a boundary condition on the IR brane, rather than by additional strong dynamics. The easiest way to accommodate this is with an odd-parity bulk scalar. However, in this case there is a competition between the light mass of the scalar relative to KK modes and the need to suppress Higgs-scalar interactions. Ref.~\cite{Matthias} suggests an intermediate boundary condition, where the scalar has some but not dominant support on the IR brane, to accommodate both constraints. However, the severity of the Higgs-scalar constraint most likely argues for a scalar mass arising both directly, and from self-interactions with some cancellation among the various contributions.

\begin{figure}[htb]
\begin{center}
\includegraphics[width=10cm]{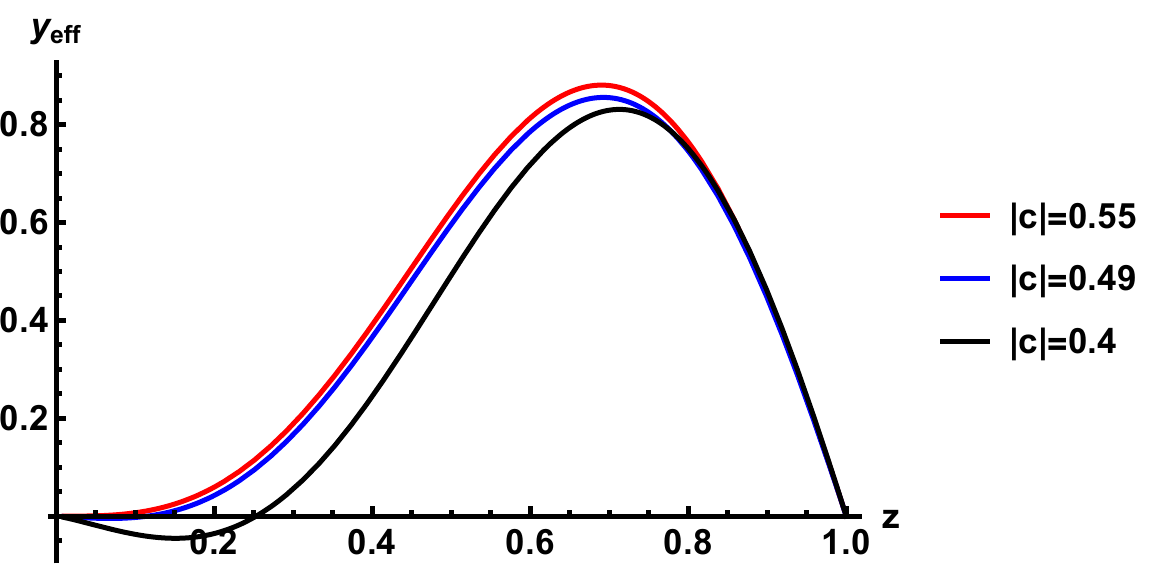}
\end{center}
\caption{The  effective Yukawa coupling between the singlet $S$ and the LH and RH modes of the first KK fermion.} \label{fig:effectiveYukawa}
\end{figure}

\subsection{CFT Interpretation\label{sec:CFT}}

As is true for RS scenarios, the model presented here has a dual CFT interpretation. There are two plausible explanations for the additional scalar localized in the bulk away from the IR brane, which most likely depends on how close the scalar is to the IR brane. One option is that there is an additional interaction which becomes strong at scale well above a TeV. In this case the two dynamical scales might be unrelated, so without an unexpected coincidence of scales,  the S-brane would be  deeper in the bulk. 

Another perhaps more attractive possibility, well-suited to a scalar localized close to the IR brane as would be true for the resonance suggested at the LHC,  is that the S-brane corresponds to states in the CFT that become massive and are integrated out at energies just above the IR brane. In the 4D language these could be some fermions of the CFT with a small mass term that breaks conformality. Once these fermions are integrated out, the $\beta$-function becomes non-zero, and the theory quickly confines and gives rise to the actual IR brane at the TeV scale. Such states would correspoond to bulk scalars localized on a tensionless brane,  as is the case with the model considered here. This mechanism would explain the proximity of the S-brane to the TeV brane. 

In this language the interpretation of the S-brane is clear: it is a domain wall inside the AdS space, which is separating an AdS region corresponding to the truly conformal part of the theory from a non-AdS region corresponding to the confining dynamics of a strongly coupled theory. We are approximating this setup with a tensionless brane in AdS space, which should suffice for exploring quantities not related to the dynamics of the confinement mechanism. However a full simulation would incorporate the modification of the background away from AdS as in~\cite{braneless}.

\section{Diphoton resonance phenomenology}

Given the setup above we are now ready to describe the phenomenology of $S$. Since it couples only to fermions, one chirality of which vanishes in the IR, the dominant couplings to SM fields will be the loop-induced couplings to gauge bosons. We will be using the notation of~\cite{Rattazzi} with the effective operators defined as 
\begin{equation}
\frac{g^2}{2\Lambda_W} S W_{\mu\nu}^2 + \frac{{g'}^2}{2\Lambda_B} S B_{\mu\nu}^2+ \frac{g_s^2}{2\Lambda_g} S G^2
\end{equation}
The effective photon coupling 
 \begin{equation}
\frac{e^2}{2\Lambda_\gamma} S F^2 
\end{equation}
is obtained by $1/\Lambda_\gamma^2 = 1/\Lambda_W^2 +1/\Lambda_B^2$. 

The standard fermion triangle loops give the contributions~\cite{HiggsHunters} 
\begin{eqnarray}
\frac{1}{\Lambda_B}= \frac{1}{8\pi^2 M_S} \sum_i y_i d_i Y_i^2 \frac{1}{\sqrt{\tau_i}} F_{\frac{1}{2}} (\tau_i ) \\
\frac{1}{\Lambda_{g,W}}= \frac{1}{8\pi^2 M_S} \sum_i y_i d_i \mu_i \frac{1}{\sqrt{\tau_i}} F_{\frac{1}{2}} (\tau_i )
\end{eqnarray}
where $Y_i$ is the hypercharge of the fermion, $y_i$ is the relevant Yukawa coupling, $d_i$ is the dimensionality of the representation, and $\mu_i$ is the Dynkin index normalized to 1/2 for a fundamental fermion, and the variable $\tau_i = \frac{4 M_i^2}{M_S^2}$. $F_{1/2}$ is the standard fermion triangle function  $F_{1/2} (\tau )= -2\tau \left[ 1+(1-\tau ) \arcsin^2\sqrt{\frac{1}{\tau}} \right] $ which for $M_i\gg M_S$ will converge very quickly 
$ F_{1/2} \to -\frac{4}{3}$. In this limit the expressions of the resonance couplings to gauge bosons induced by the first KK mode is given by 
\begin{eqnarray} 
 \frac{1}{\Lambda_B} &=& \frac{1}{4\pi^2 M_{KK}} \left[ \frac{1}{6} y_Q +\frac{4}{3} y_u + \frac{1}{3} y_d + \frac{1}{2} y_L +y_e \right]  \\
\frac{1}{\Lambda_W} &=& \frac{1}{8\pi^2 M_{KK}} \left[  3 y_Q+ y_L  \right]  \\
\frac{1}{\Lambda_\gamma} &=& \frac{1}{4\pi^2 M_{KK}} \left[ \frac{5}{3} y_Q +\frac{4}{3} y_u + \frac{1}{3} y_d + y_L +y_e \right] \\
\frac{1}{\Lambda_g} &=& \frac{1}{8\pi^2 M_{KK}} \left[  2 y_Q+ y_u +y_d \right]  \\
\end{eqnarray}
 where we have assumed degenerate KK fermions, and flavor-independent Yukawa couplings to the resonance $S$ for each type of SM fermion.

\begin{figure}[htb]
\begin{center}
\includegraphics[width=4cm]{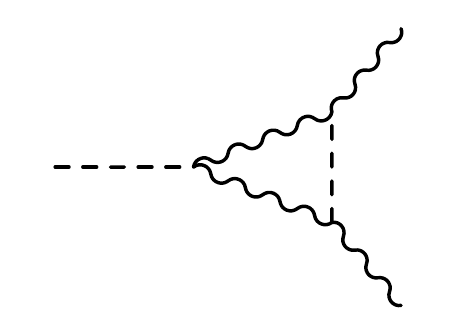}
\end{center}
	    \caption{Correction to the effective $SF^2$ operator used for the NDA estimate for $y_{{\rm NDA}}$.\label{fig:NDA}}
\end{figure}

 These results are in nice agreement with the expected NDA estimates. A generic $y S  F^2$ coupling generates  a quadratically divergent correction (see Fig.~\ref{fig:NDA}):
\begin{equation}
\Delta y \sim y^3 \frac{\Lambda^2}{16\pi^2}
\end{equation}
giving rise to $y_{{\rm NDA}} \sim \frac{4\pi}{\Lambda}$. For the cutoff we  use the local warped down version of the 5D cutoff $\Lambda= \frac{24 \pi^3}{g_5^2} \frac{R}{z}$, which together with the matching of the couplings wgives $y_{{\rm NDA}} \sim \frac{g_4^2}{6\pi^2} \frac{x_1}{m_{KK}} z_0 \log \frac{R'}{R}$. Note that the perturbative answer from the first KK mode is almost exactly the NDA answer (while of course  NDA is in the limit of strong coupling). Because our result involves perturbative Standard Model gauge couplings, the results differ by the explict couplings in the exact result. Furthermore, the perturbative calculation with only the first KK mode doesn't contain the warp factor.  For models where the brane is deeper in the bulk, where a heavier KK mode couples more strongly,  the warp factor suppression expressed here via $z_0$ would be present.

Using the expressions for the decay widths~\cite{Rattazzi,Gilad,Tomer,Michele}
\begin{equation}
\Gamma (S\to \gamma \gamma) = \pi \alpha^2 \frac{M_S^3}{\Lambda_\gamma^2} \ , \ \Gamma (S\to  gg) = \pi \alpha^2_s \frac{M_S^3}{\Lambda_g^2} 
\end{equation}
and assuming for simplicity one overall effective Yukawa coupling, the decay width to photons is 
\begin{equation}
\Gamma_\gamma \approx \alpha^2 \frac{16 y_{eff}^2}{9\pi^3} \frac{M_S^3}{M_{KK}^2} \sim   2 \cdot 10^{-4} \ {\rm GeV} \cdot y_{eff}^2.
\end{equation}
For the numerical value we have used $M_{KK}\sim 2.5$ TeV. This corresponds approximately to the bound on the KK gluon mass in bulk RS models, which in the simplest versions are expected to be approximately degenerate with the KK fermions in the absence of additional brane-localized terms. 

The relative branching ratio to photons (vs. gluons) is  
\begin{equation}
{\rm Br} (S\to \gamma\gamma) = \frac{\frac{\alpha^2}{\Lambda_\gamma^2}}{\frac{\alpha^2}{\Lambda_\gamma^2}+\frac{8\alpha_s^2}{\Lambda_g^2}} \sim 5 \cdot 10^{-3}\ .
\end{equation}
Thus $S$ is a narrow resonance with a width of order 40 MeV mainly decaying to gluons. The total production cross section of the resonance dominated by gluon fusion would be $\sigma_{gg\to S} \sim \ {\rm pb}$ leading to ${\cal O}(3000)$ events in each experiment, of which about 15 would correspond to signal diphoton events, in agreement with the observed rates in ATLAS and CMS. 

Finally, the charges and multiplicities of the fermions KK modes will determine the relative ratios for decays to $WW,ZZ$ and $Z\gamma$. Assuming universal Yukawa coupling factors these ratios are uniquely predicted, and are given by 
\begin{eqnarray}
\frac{\Gamma_W}{\Gamma_\gamma} &=& \frac{2}{s^4} \frac{\Lambda_\gamma^2}{\Lambda_W^2} \sim 5.3 \nonumber \\
\frac{\Gamma_Z}{\Gamma_\gamma} &=&  \Lambda_\gamma^2 \left(\frac{\tan\theta_W^2}{\Lambda_B}+\frac{\cot\theta_W^2}{\Lambda_W} \right)^2 \sim 2 \nonumber \\
\frac{\Gamma_{\gamma Z}}{\Gamma_\gamma} &=&  2 \Lambda_\gamma^2 \left(\frac{\tan\theta_W}{\Lambda_B}-\frac{\cot\theta_W}{\Lambda_W} \right)^2 \sim 0.24 \nonumber \\
\end{eqnarray}
all in accordance with the current experimental constraints.

We also consider the decays $S\to t\bar{t}$ and $S\to b\bar{b}$. After EWSB there will be two sources for these. One is the $S-h$ mixing, where the $S$ inherits the couplings of the SM Higgs. However this mixing is suppressed by $(m_h/M_S)^2\sim 0.03$. The other source is the mixing between the $t_R$ and the KK top, yielding a tree-level $St\bar{t}$ Yukawa coupling which is suppressed by $m_t/M_{KK}\sim 0.07$.  With this as the only suppression factor the decay width is $0.2 y^2_{eff}$ GeV, which is abount the experimental limit for top resonance searches~\cite{Gilad}. Note however that the same mixing is responsible for a potentially large shift in the $Zb\bar{b}$ coupling, which is constrained by electroweak precision measurements at the $3\cdot 10^{-3}$ level. If wave function suppression for $(t,b)_L$ is used to reduce the corrections to $Zb\bar{b}$ then the $S\to t\bar{t}$ rate will also sharply decrease, and the decay width will be at most few$\cdot 10^{-4}$ GeV. The more common approach is the introduction of a custodial protection for the $Zb\bar{b}$ copling~\cite{custodial}, which will protect the bottom quarks from large mixings, but not the up-type quarks. This will correspond to the case considered above, where the only suppression of the Yukawa is from the mixing. 
 Consequently in such custodial models $S$ should show up in top resonance searches reasonably soon. 

The coupling (\ref{eq:SYukawa}) also contributes to the 3-body decays $S\to t\bar{t}h$ and $t\bar{t}Z$ via off-shell KK-tops. The expected rate is comparable to that of diphotons, yielding an 8 TeV $t\bar{t}h, t\bar{t}Z$ cross sections of ${\cal O}(fb)$ (compared to the 86 fb $t\bar{t}h$ associated production cross section and 150 fb $t\bar{t}Z$ cross section). While this will be a small correction to the overall $t\bar{t}h$ and $t\bar{t}Z$ cross sections, resonance searches in these channels may be more sensitive to these 3-body decay modes.

\section{Suppression of the Higgs coupling}

Most theories of a weak scale scalar subject to strong dynamics would also produce an operator $\frac{S}{f} |D_\mu H|^2$.\footnote{If neither the Higgs nor $S$ are pseudo-Goldstone bosons the $S|H|^2$ operator is also expected to be generated.}
Ref~\cite{Rattazzi} argued that for such a model to work consistent with the suppression of this operator, the couplings to electroweak gauge bosons would have to be two
orders of magnitude bigger than naive NDA estimates (including the  less naive estimates based on explicitly integrating out vector fermions). Of course, this assumes that the Higgs and scalar participated in common strong
dynamics. However there would be no explanation for the similarity of the mass of the new resonance to that of the Higgs boson.

An alternative  solution~\cite{Rattazzi} to this quandary is for the scalar to in fact be a pseudoscalar so that the dangerous operators are forbidden. Ref.~\cite{Giacomo} explores a large class of models of this sort, and argues that only a few are consistent
with all the required decay constraints. 
A different solution based on the custodial SU(2)$_R$ symmetry was presented in~\cite{toappear}. We now consider how our model might address this issue.

\begin{figure}[htb]
\begin{center}
\includegraphics[width=6cm]{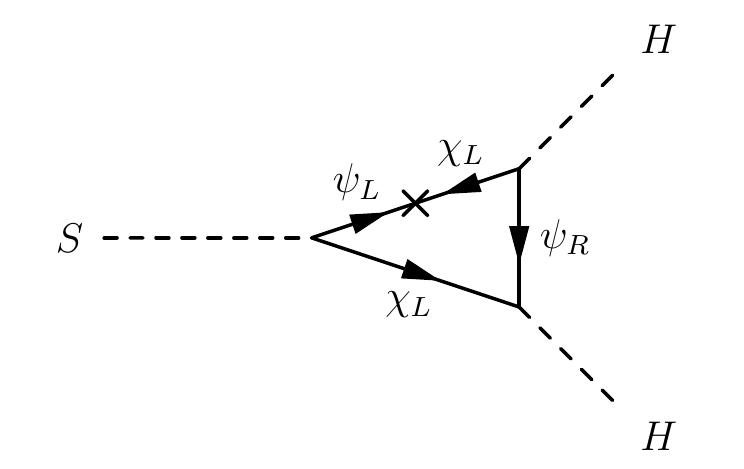}
\end{center}
	    \caption{An example of a contribution to the induced $S|H|^2$ or $S|D_\mu H|^2$ operators from fermion KK modes. The cross indicates an insertion of the KK mass.\label{fig:SHHloop}}
\end{figure}

Currently the decay rate of a new scalar to Higgs bosons is bounded to be less than about 30 times the branching to photons~\cite{Gilad}.  As with decays to photons, loop-induced diagrams (see for example Fig.~\ref{fig:SHHloop}) can mediate $S\to hh$ via the operators $y m_{KK} S |H|^2$ or $\frac{y}{m_{KK}} S |D_\mu H|^2$, with $y\sim y_{eff}^2 Y^4\frac{24}{16\pi^2}$, where the 24 is  the multiplicity of the KK fermion doublets running in the loop, $y_{eff}$ is the effective Yukawa coupling of $S$ to the KK fermions (\ref{eq:SYukawa}), while $Y$ is the dimensionless Yukawa coupling of the Higgs to the KK modes, expected to be an ${\cal O} (1)$ number.

\begin{figure}[htb]
\begin{center}
\includegraphics[width=9cm]{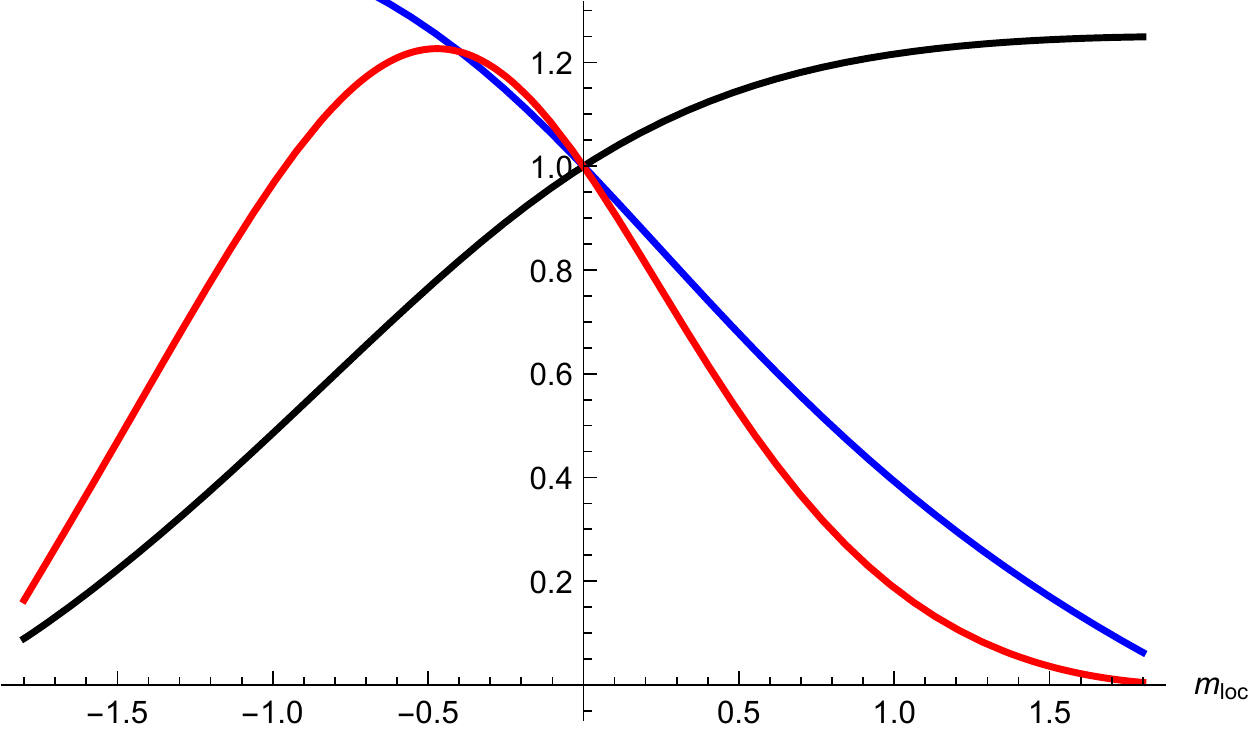}  
\end{center}
\caption{The dependence of the KK mass (black) and the IR brane wave function (blue) on the localized fermion mass on the S-brane. Both curves are ratios normalized to the case with a vanishing brane localized mass. The red curve is the product of the KK mass ratio and the square of the IR brane wave function, the quantity whose fourth power will eventually determine the decay width to Higgses.    \label{fig:mloc1}}
\end{figure}

\begin{figure}[htb]
\begin{center}
\includegraphics[width=8cm]{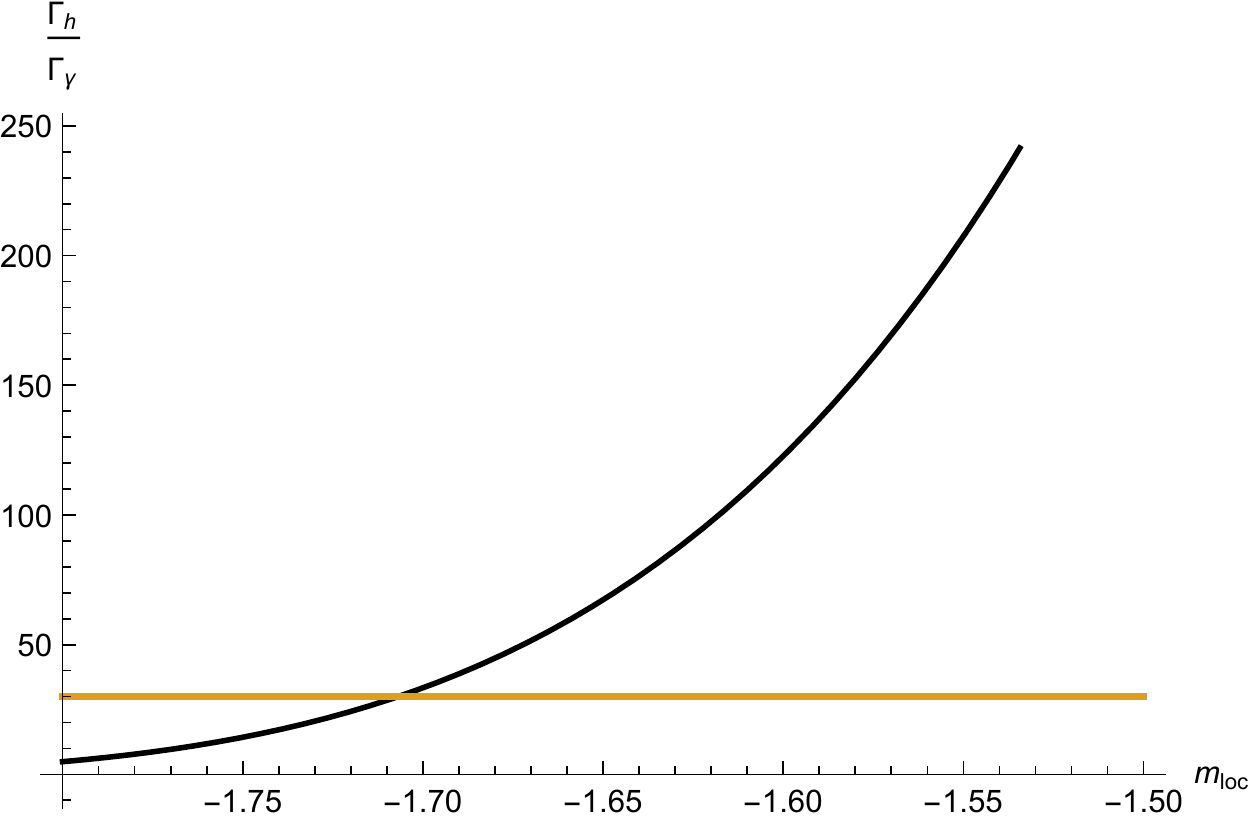} \hspace*{1cm}   \includegraphics[width=8cm]{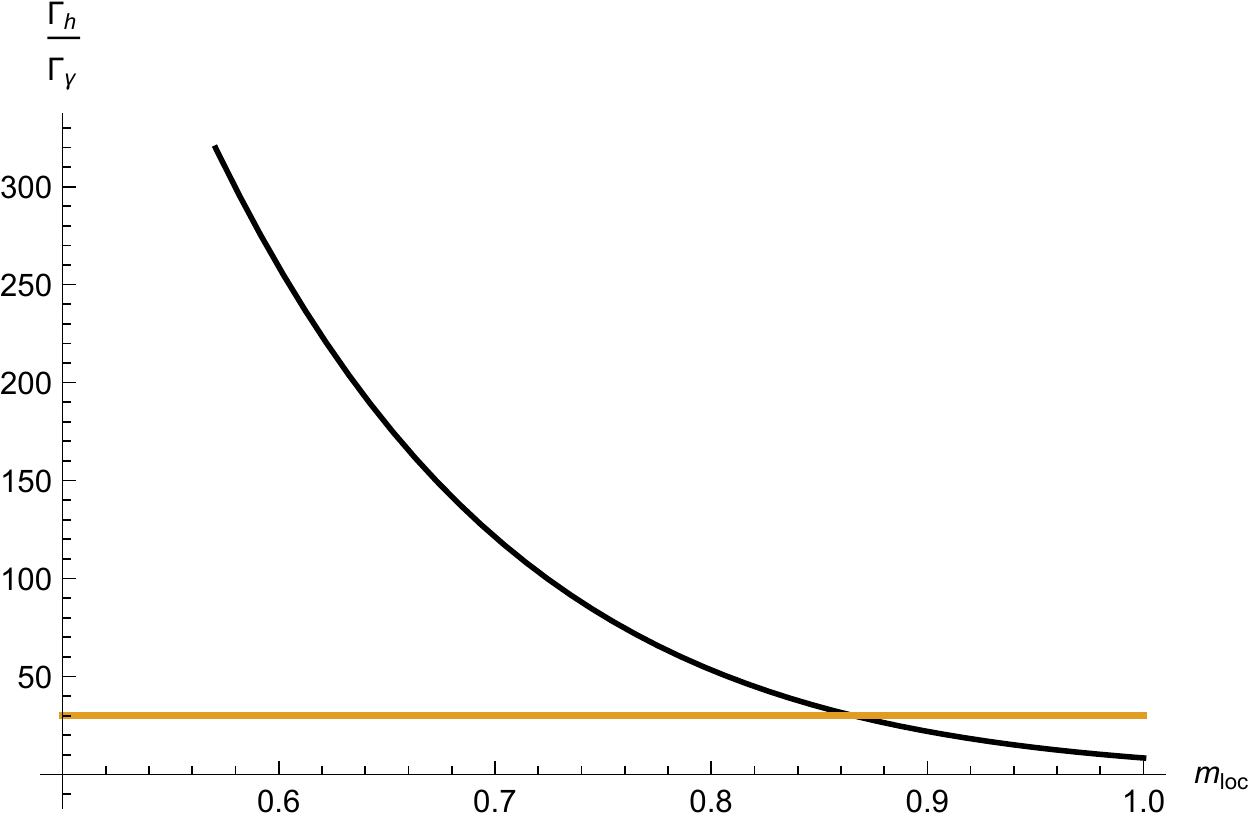}
\end{center}
\caption{The ratio of the $S\to hh$ and $S\to \gamma\gamma$ decay rates as a function of a localized fermion mass on the S-brane. Here we assumed the position of the S-brane at $0.8 R'$ and a common bulk fermion mass of $c=0.55$. The horizontal line indicates the experimental bound. The left plot is a zoom-in to the region with negative localized masses, where the suppression of the KK mass will lead to sufficiently low decay rates to Higgses. The right plot is the zoom-in to the region with a positive localized fermion mass, where the suppression of the wave function will dominate. 
 \label{fig:mloc2}}
\end{figure}

 The resulting ratio of decay widths to Higgses vs photons for the case when the $S|H|^2$ operator is allowed is
\begin{equation} 
\frac{\Gamma_{S\to hh}}{\Gamma_{S\to\gamma\gamma}}  \sim 65 \left(\frac{M_{KK}}{M_S}\right)^4 Y^4,  
\end{equation}
while for the case when the Higgs is a pseudo-Goldstone boson (corresponding to the $S|D_\mu H|^2$ operator) it is 
\begin{equation} 
  \frac{\Gamma_{S\to hh}}{\Gamma_{S\to\gamma\gamma}}  \sim 65\,  Y^4\ . 
\end{equation}
The pseudo-Goldstone boson Higgs~\cite{PGBHiggs} case needs a mild suppression from the effective Yukawa couplings, while the non-Goldstone case is well outside the experimental bound without additional suppression for the values of the KK mass we have assumed. However, the extra-dimensional scenario allows for a lower fermionic KK mass when there is a mass term localized to the S brane.   Such a mass is expected to be generated naturally due to a tadpole induced VEV for $S$: for example if the sign of the induced mass cancels the KK mass for the bulk fermions the first KK mass can be lowered all the way to the experimental bound of around 700 GeV, removing the $\left(\frac{M_{KK}}{M_S}\right)^4$ enhancement for the non-Goldstone Higgs case. For the opposite sign of the localized mass the KK mass increases slightly, however the fermion wave function will be suppressed on the IR brane, leading to sequestering of the $S$ from the Higgs.  In Fig.~\ref{fig:mloc1} we show the effect of a localized fermion mass on the S-brane on the KK spectrum as well as the fermion wave function on the 
IR brane, while in Fig.~\ref{fig:mloc2} we combine these two effects and show the full dependence of the ratio of widths  $\frac{\Gamma_{S\to hh}}{\Gamma_{S\to\gamma\gamma}} $ on the localized mass. We can see that for a sufficiently large localized mass greater than 0.81 or less than -1.71 the rates of $S\to hh$ decays reduce below the experimental bound.  

\section{Conclusions}

We have shown why a warped AdS geometry in the context of bulk RS with an additional bulk-localized scalar has all the ingredients necessary to account for the hinted-at LHC resonance. The large number of KK fermion resonances gives rise to a sufficiently large interaction of the scalar with two gauge bosons with all Yukawas order unity. The sequestering of the scalar from the IR brane, which was necessary for the Yukawas given the projected chiral modes on the IR brane and the consequent small wave function for one chirality of KK mode, suppresses the interaction between the scalar and the Higgs boson at tree-level.

 We note that bulk RS provides a good template in which to address flavor when the SM fermions correpond to bulk fields. In all such models, there are many KK modes but our result depends only weakly on the detailed flavor model as the bulk KK modes depend only slightly on the bulk fermion mass parameters that determine the fermion mass hierarchy. The results we have presented are  in fact conservative in that we assume most fermion zero modes are concentrated on the UV brane in order to explain their small interaction with the Higgs boson. If localized on the IR brane, the interaction with the new scalar would be slightly bigger and the interaction of S with Higgs bosons mediated by KK modes would be significantly smaller.

 Furthermore, the decays to weak gauge bosons are naturally suppressed to the necessary level to be consistent with their non-observation with no tweaking required for the ratio of Yukawas.   Detailed Yukawa ratios determine the precise branching fractions predicted. But it is clear that decays to weak gauge bosons are generically consistent with existing bounds. Furthermore, decays to jet pairs are well within the experimental constraint when all Yukawas are of order unity. This prediction too can change with varying Yukawa couplings of the different fermion types. 

With Yukawas all of order unity the  resonance will be narrow with a total width of ${\cal O}(50 \ {\rm MeV})$. This can be increased if the Yukawa couplings of the quarks are bigger, but a large boost is unlikely within the perturbative regime. 

 Turning around our results, even independently of the hinted-at resonance, searches of this kind are a good way to look for bulk RS, which could be notoriously challenging to find. Gauge bosons are expected to be the dominant decay modes of bulk scalars since other fields are likely to be localized in the IR or UV. From the dual CFT point of view this corresponds to gauge invariance protecting the dimension of the gauge bosons, whereas other operators have dominant support in the IR or UV. Future measurements with more data will extend the effective KK reach considerably. Of course, this model has the assumption of an additional localized scalar. But given that the model with heavier KK modes might be inaccessible, this additional search can be a useful supplement.

\section*{Acknowledgments}

We thank Kiel Howe, Eric Kuflik, Markus Luty, Matt Reece and Daniel Stolarsky for useful discussions, and to Martin Bauer and Matthias Neubert on discussions on the loop induced $S\to hh$ decays.   While this work was nearing completion Ref.~\cite{Matthias} appeared which is considering a similar model.  C.C.~is supported in part by the NSF grant PHY-1316222.  LR is supported in part by the NSF grant PHY-1415548.




\end{document}